\documentclass[
    ,final            
  ]
  {aipproc}

\layoutstyle{6x9}

\begin{document}
\title{Asymmetric Line Profiles in Spectra of Gaseous Metal Disks Around Single White Dwarfs}
\classification{97.82.Jw, 97.20.Pp, 97.10.Gz}
\keywords      {stars: individual (SDSS\,J122859.93+104032.9) -- white dwarfs -- circumstellar matter -- planetary systems}
\author{Stephan~Hartmann}{
  address={Institute for Astronomy and Astrophysics,
    Kepler Center for Astro and Particle Physics,\\
    Eberhard Karls University,
    D-72076 T\"ubingen,
    Germany}}
\author{Thorsten~Nagel}{
  address={Institute for Astronomy and Astrophysics,
    Kepler Center for Astro and Particle Physics,\\
    Eberhard Karls University,
    D-72076 T\"ubingen,
    Germany}}
\author{Thomas~Rauch}{
  address={Institute for Astronomy and Astrophysics,
    Kepler Center for Astro and Particle Physics,\\
    Eberhard Karls University,
    D-72076 T\"ubingen,
    Germany}}
\author{Klaus~Werner}{
  address={Institute for Astronomy and Astrophysics,
    Kepler Center for Astro and Particle Physics,\\
    Eberhard Karls University,
    D-72076 T\"ubingen,
    Germany}}
\begin{abstract}
Metal-rich debris disks were discovered around several single DAZ and DBZ white dwarfs, which may stem from disruption of smaller rocky planetesimals. In some cases, the material in addition forms a gaseous disk. For SDSS\,J122859.93+104032.9, the double peaked infrared Ca\,\textsc{ii} triplet at $\lambda\lambda$\,8498,\,8542,\,8662\,{\AA} exhibits a strong red/violet asymmetry. With the T\"ubingen non-LTE \emph{Ac}cretion \emph{D}isk \emph{c}ode \textsc{AcDc} we calculated the spectrum and vertical structure of the disk, assuming the chemical mixture of the disk's material being similar to a chondrite-like asteroid, the most prominent type in our own solar system. By modifying the code to simulate different non axis-symmetrical disk geometries, the first preliminary results are in good agreement with the observed asymmetric line profile.
\end{abstract}

\maketitle
%
%
\section{Motivation}
\subsection{Metal-Rich Dust Disks}
Starting in 1987 with the analysis of G\,29-38 by \citet{1987Natur.330..138Z}, about twenty metal-enriched white dwarfs with a significant infrared excess in their spectra have been discovered, without a cool companion being found. Due to their high surface gravity the sedimentation time of these objects is rather short \citep{1997A&A...320L..57K}. To explain the high metallicity by ongoing accretion, a metal-rich dust cloud was suggested as a donator source. Spitzer observations confirmed the dust material \citep{2005ApJ...635L.161R}, which is located in an equatorial plane \citep{1990ApJ...357..216G} forming a metal-rich but hydrogen- and helium-poor \citep{2003ApJ...584L..91J} dust disk.
\subsection{Metal-Rich Gas Disks}
\begin{figure}%
  \centering%
  \includegraphics[width=0.99\textwidth]{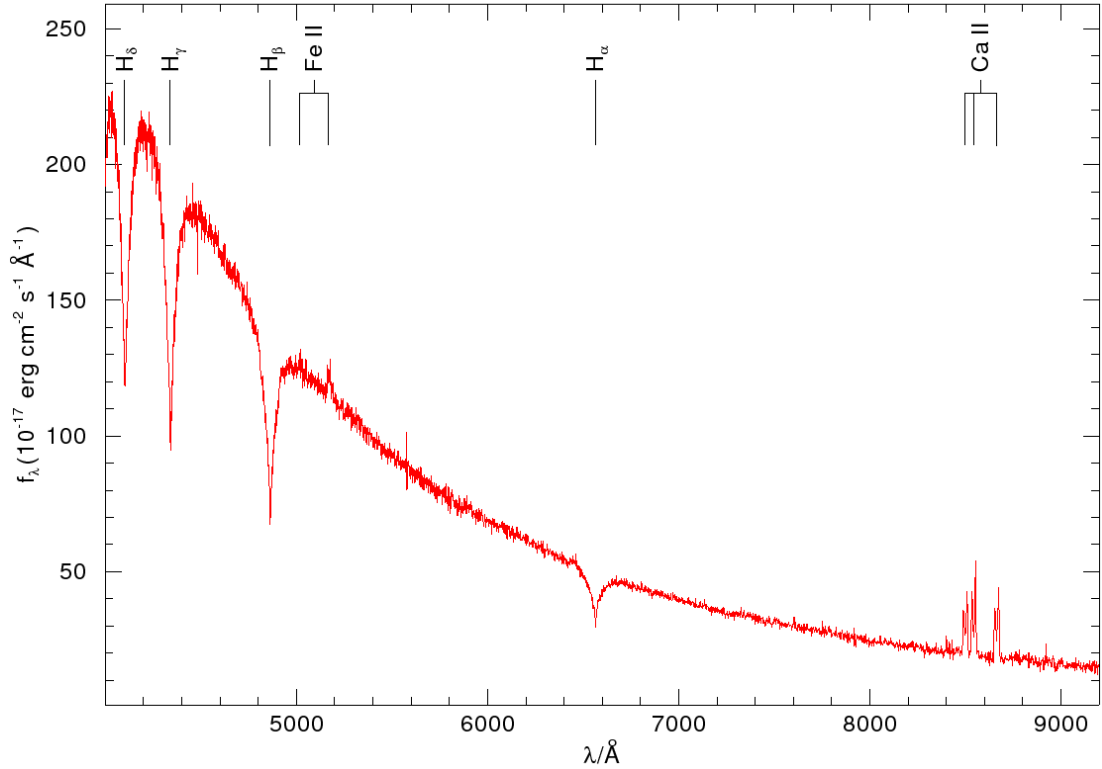}%
  \caption{Spectrum of SDSS\,J1228+1040 from the Sloan Digital Sky Survey.}%
  \label{fig:1}%
\end{figure}

For the hottest of those white dwarfs, with $T_{\mathrm{eff}}\ge 20\,000\,\mathrm{K}$, \citet{2006ApJ...646..474K} proclaimed that their dust material is sublimated to a gaseous disk. And indeed, five white dwarfs, found in the Sloan Digital Sky Survey (e.g. Fig.\,\ref{fig:1}) by \citet[priv.~comm.]{2006Sci...314.1908G, 2007MNRAS.380L..35G, 2008MNRAS.391L.103G} indicate gas disk emission features in their spectra in addition to the IR spectral component of a dust disk. These gas features are mainly the Ca\,\textsc{ii} $\lambda\lambda$\,8498,\,8542,\,8662\,{\AA}, although at least two of the five spectra also show a Fe\,\textsc{ii} $\lambda\lambda$\,5018,\,5169\,{\AA} in emission.
\subsection{Origin of the Disk Material}
Former planetary systems around the white dwarfs progenitor may survive the host stars' late evolutionary phases, as different theoretical approaches have shown (e.g. \citet{2008AJ....135.1785J}). After the actual formation of the white dwarf, a smaller rocky body, an asteroid or planetesimal, might have been disturbed in its orbital motion by another surviving, larger planet. When it reaches the inner part of the system close to the central object, it will get tidally disrupted, providing the material to form the dust and gaseous disk.
\section{Aim and Method}
\subsection{Asymmetric Line Profiles}
\begin{figure}%
  \centering%
  \includegraphics[width=0.8\textwidth]{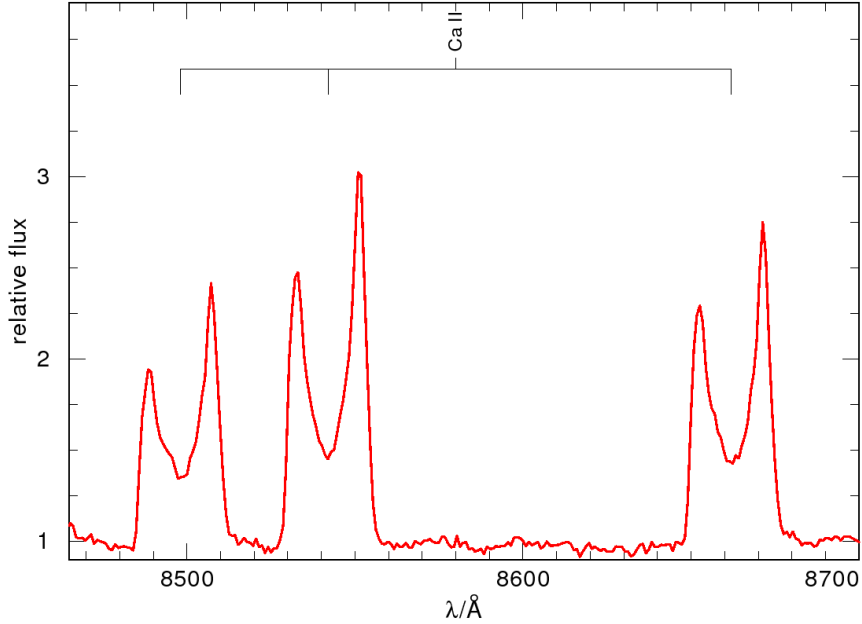}%
  \caption{Ca\,\textsc{ii} $\lambda\lambda$\,8498,\,8542,\,8662\,{\AA} in the spectrum of SDSS\,J1228+1040. Each line is split due to the Kepler rotation of the disk but shows an asymmetry in the relative strengths of the red/violet peaks, favoring the red wing peak in this case.}%
  \label{fig:2}%
\end{figure}

The disk's spectral line features are expected to be double peaked as the emitting material rotates with its Kepler
velocity around the white dwarf. Surprisingly, these double-peak structures show a significant difference in the
red/violet peaks line strengths, shown in Fig.\,\ref{fig:2} for the prominent Ca\,\textsc{ii} feature of
SDSS\,J122859.93+104032.9 (hereafter SDSS\,J1228+1040). We assumed this asymmetry to directly originate from an actual
asymmetric disk geometry. Hydrodynamically simulating the shape of a disk changing with time, we later modified our spectrum-simulation code according to these geometrical situations.
\subsection{Hydrodynamic Simulation with \textsc{fargo}}
For the hydrodynamic investigation of a viscous gas-disk we used the code \textsc{fargo} (\emph{F}ast \emph{A}dvection in \emph{R}otating \emph{G}aseous \emph{O}bjects) by \citet{2000A&AS..141..165M}. Even though being almost exclusively used in the calculation of protoplanetary disks and planet formation scenarios, this code is suited for all kinds of sheared fluid disks. We calculated the temporal evolution of a blob of material, starting with a Gaussian density distribution (parameters given in Tab.\,\ref{tab:1}).\par
\begin{figure}[!Ht]%
\parbox{\textwidth}{
\begin{center}
    {\includegraphics[width=0.65\textwidth]{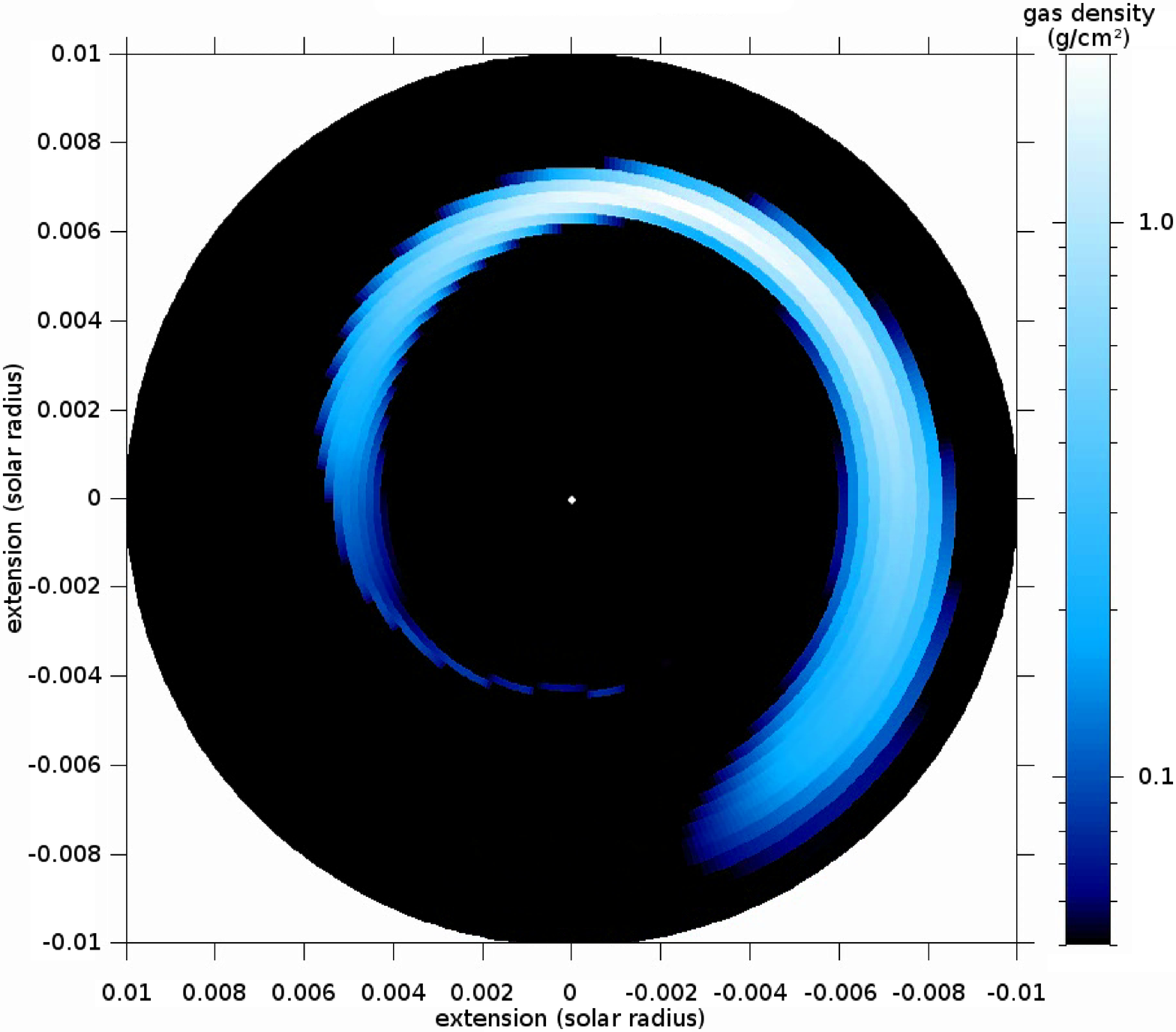}}
    {\includegraphics[width=0.65\textwidth]{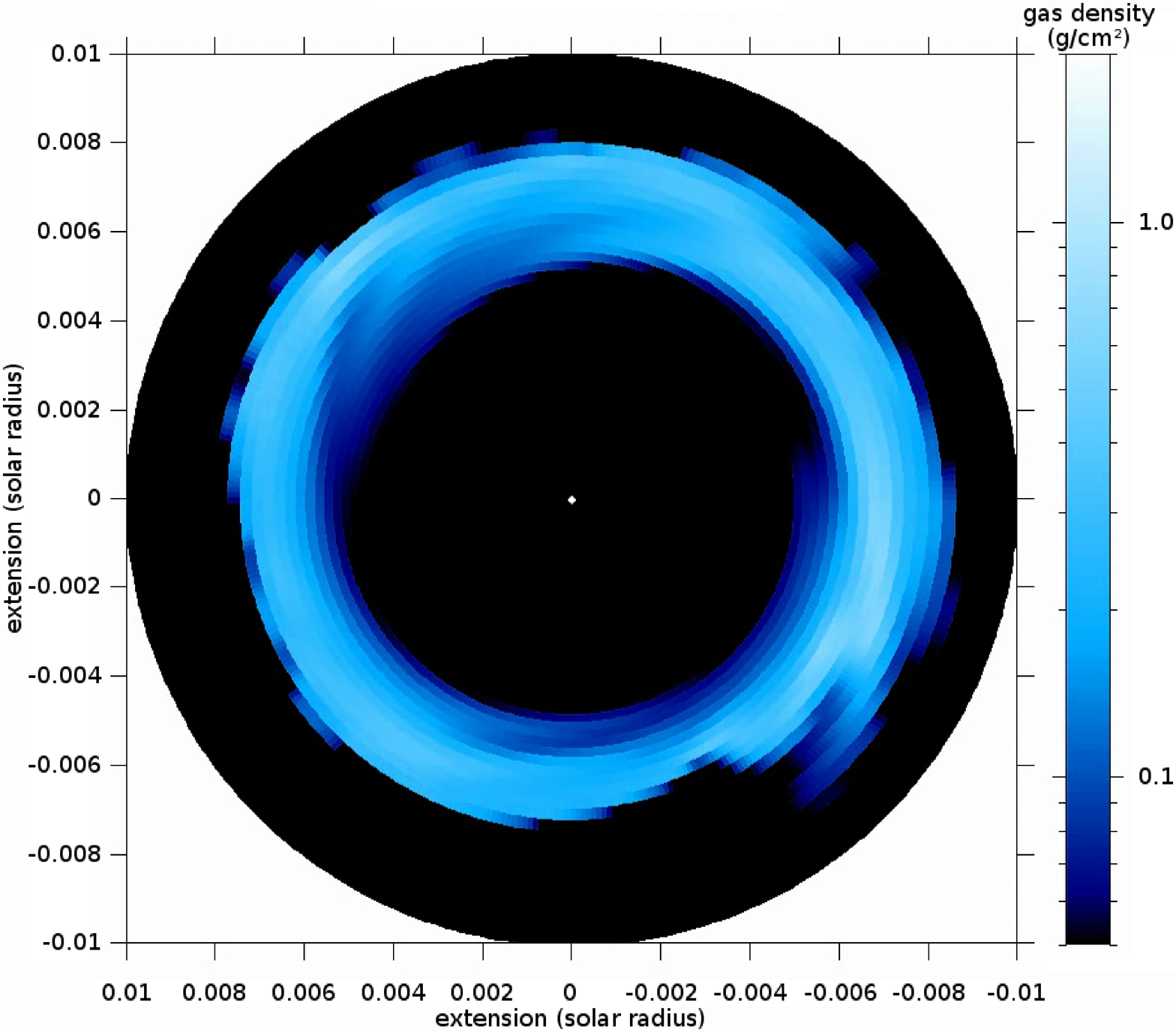}}
\end{center}
    \caption{The two selected evolution phases of the \textsc{fargo} simulation -- color-coded is the surface mass density on a logarithmic scale. The top panel shows an early stage $\left(t_{\mathrm{sim}}\approx 40\,\mathrm{yr}\right)$ as the disks forms a spiral-arm like structure. The lower panel shows a late $\left(t_{\mathrm{sim}}\approx 125\,\mathrm{yr}\right)$, but with several hundreds of orbits, long-lasting phase of a fully closed disk with an out-of-balance matter distribution.}%
    \label{fig:4}%
}
\end{figure}

\begin{table}[b]%
  \begin{tabular}{p{8em}lr}\hline
    White dwarf mass&$M_{\mathrm{WD}}/\mathrm{M}_{\odot}$&$0.77$\\
    White dwarf radius&$R_{\mathrm{WD}}/\mathrm{km}$&$7700$\\
    Disk mass&$M_{\mathrm{disk}}/\mathrm{g}$&$7\cdot 10^{21}$\\
    Disk outer radius&$R_{\mathrm{o}}/R_{\mathrm{WD}}$&$136$\\
    Simulation rim&$R_{\mathrm{sim}}/R_{\mathrm{o}}$&$1.5$\\\hline
  \end{tabular}%
  \caption{\textsc{fargo} simulation parameters}%
  \label{tab:1}%
\end{table}
To perform the modifications on the spectrum synthesis code, we selected two geometries, a spiral arm structure of the early simulation phase and a fully closed but out-of-balance disk from the late stage of the evolution (Fig.\,\ref{fig:4}).
\subsection{Spectral Synthesis Code \textsc{AcDc}} 
\begin{figure}%
\parbox{\textwidth}{
\begin{center}
    {\includegraphics[width=0.65\textwidth]{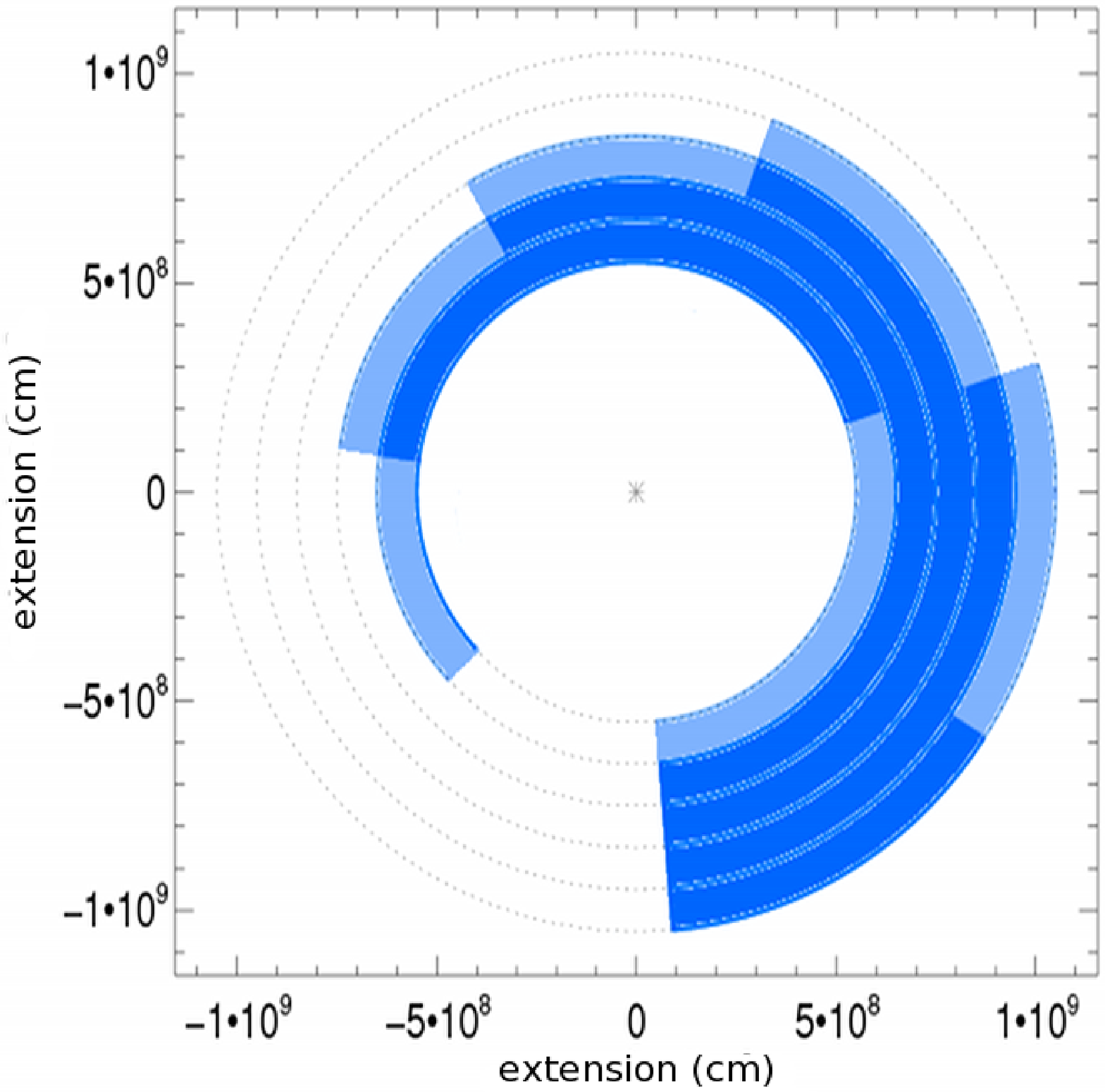}}
    {\includegraphics[width=0.65\textwidth]{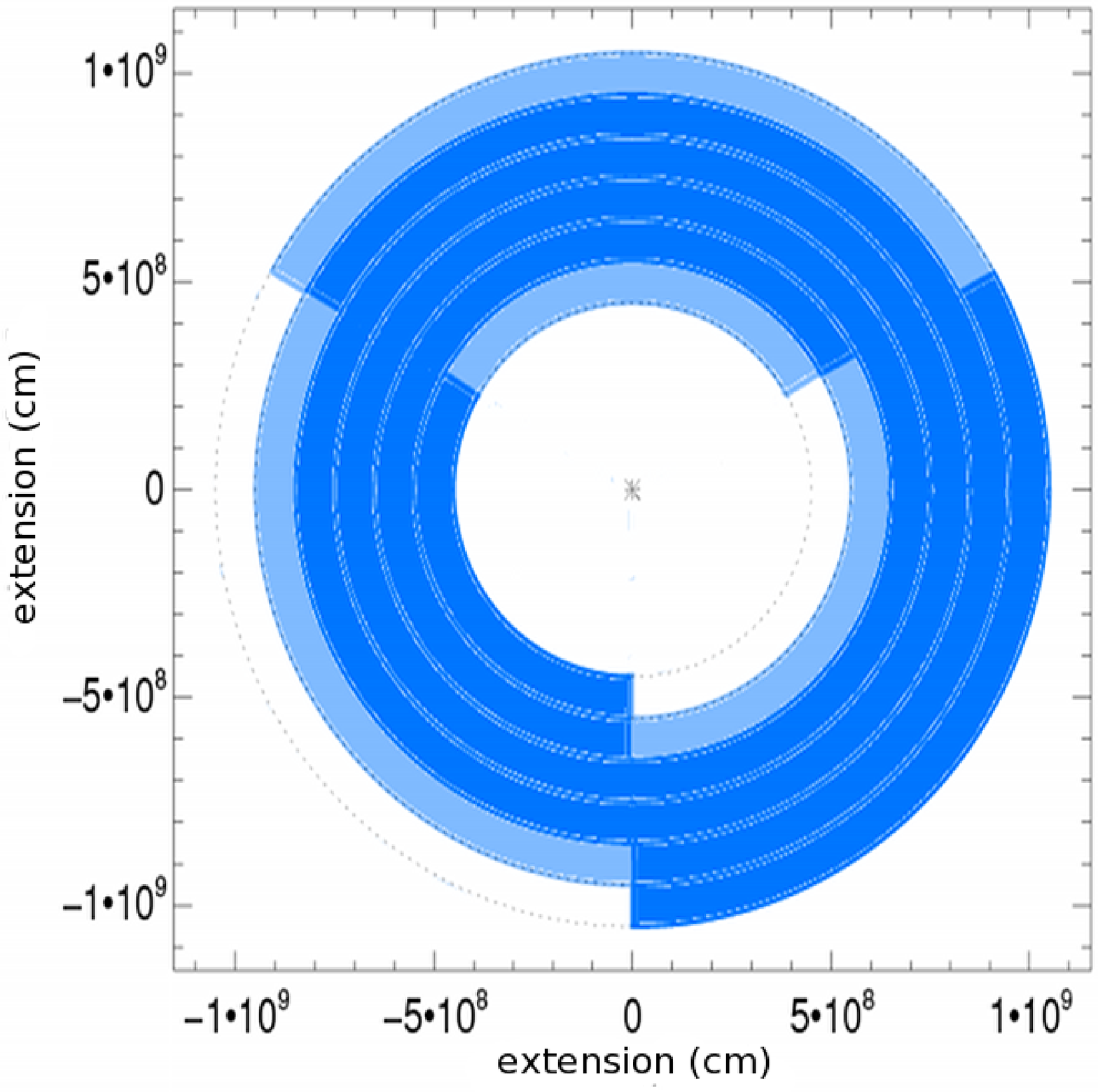}}
\end{center}
    \caption{Geometry maps representing the selected \textsc{fargo} phases shown in Fig.\,\ref{fig:4}. The two colors (greyshades) indicate that the flux of the adjoint ring segment is set to zero and \textsc{AcDc} will compute a downward interpolation. For the subsequent spectrum calculations the observer is meant to look from the lower end of the box towards the central object.}%
    \label{fig:6}%
}
\end{figure}

Using the \emph{Ac}cretion \emph{D}isk \emph{c}ode (\textsc{AcDc}) developed by \citet{2004A&A...428..109N} to calculate the disks' spectra, we assumed a thin $\alpha$-disk as described by \citet{1973A&A....24..337S}. This allows us to decouple the vertical and radial structure and separate the disk into concentric annuli. For each of these rings with radius $R$ the effective temperature $T_{\mathrm{eff}}$ as well as a viscosity times surface mass density $w\Sigma$ value is derived for given values of the white dwarf's radius $R_{\mathrm{WD}}$, mass $M_{\mathrm{WD}}$, accretion rate $\dot{M}$, the gravitational constant $G$, and the Stefan-Boltzmann constant $\sigma$ from
\begin{eqnarray}
  T_{\mathrm{eff}}\left(R\right)&=&\left[\frac{3GM_{\mathrm{WD}}\dot{M}}{8\pi\sigma R^{3}}\left(1-\sqrt{\frac{R_{\mathrm{WD}}}{R}}\right)\right]^{\frac{1}{4}}%
  \label{eq:1}\\
  w\Sigma\left(R\right)&=&\frac{\dot{M}}{3\pi}\left(1-\sqrt{\frac{R_{\mathrm{WD}}}{R}}\right)\mathrm{.}%
  \label{eq:2}
\end{eqnarray}
The program then simultaneously solves the following set of equations for each of the rings:
\begin{itemize}
\item Radiation transfer for the specific intensity $I\left(\nu,\mu\right)$,
\item Hydrostatic equilibrium between gravitation, gas and radiation pressure,
\item Energy conservation for viscously generated $E_{\mathrm{mech}}$ and radiative $E_{\mathrm{rad}}$ and
\item Static rate equations $\frac{\partial n_i}{\partial t}=0$ for the population numbers $n_i$ of the atomic level $i$ of the given model atom in the non-LTE case.
\end{itemize}
For the subsequent spectral surface integration phase of \textsc{AcDc}, we constructed simple geometry maps (Fig.\,\ref{fig:6}) following the selected \textsc{fargo} pictures. This modification allowed us to reset the spectral flux to zero for those ring segments, which are not part of the structure. By simple linear interpolation between the neighboring annuli spectra, \textsc{AcDc} integrates the spectrum of the whole non axis-symmetrical disk.
\section{Synthetic Spectra and Results}
\begin{figure}%
\parbox{\textwidth}{
\begin{center}
    {\includegraphics[width=0.85\textwidth]{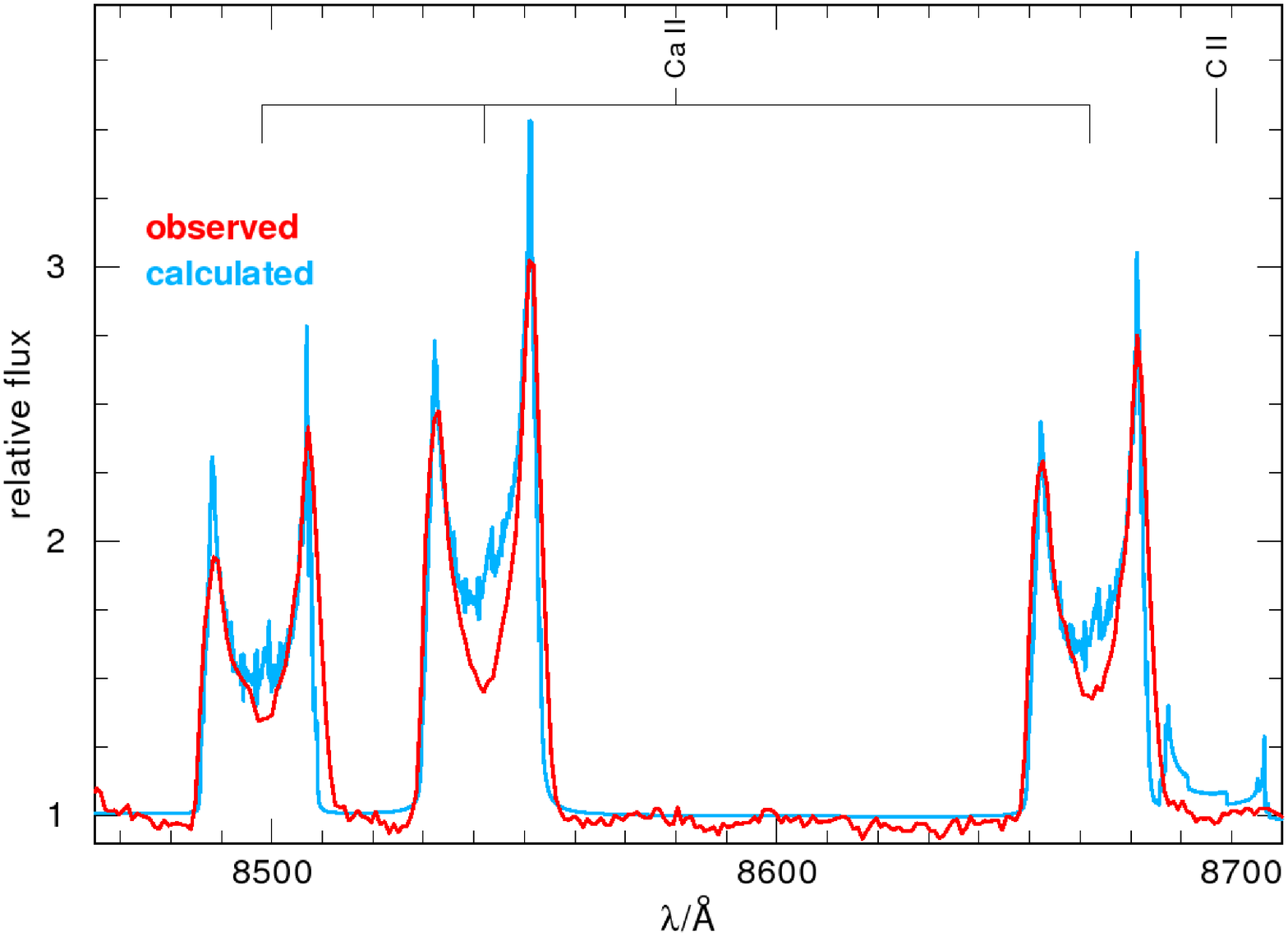}}
    {\includegraphics[width=0.85\textwidth]{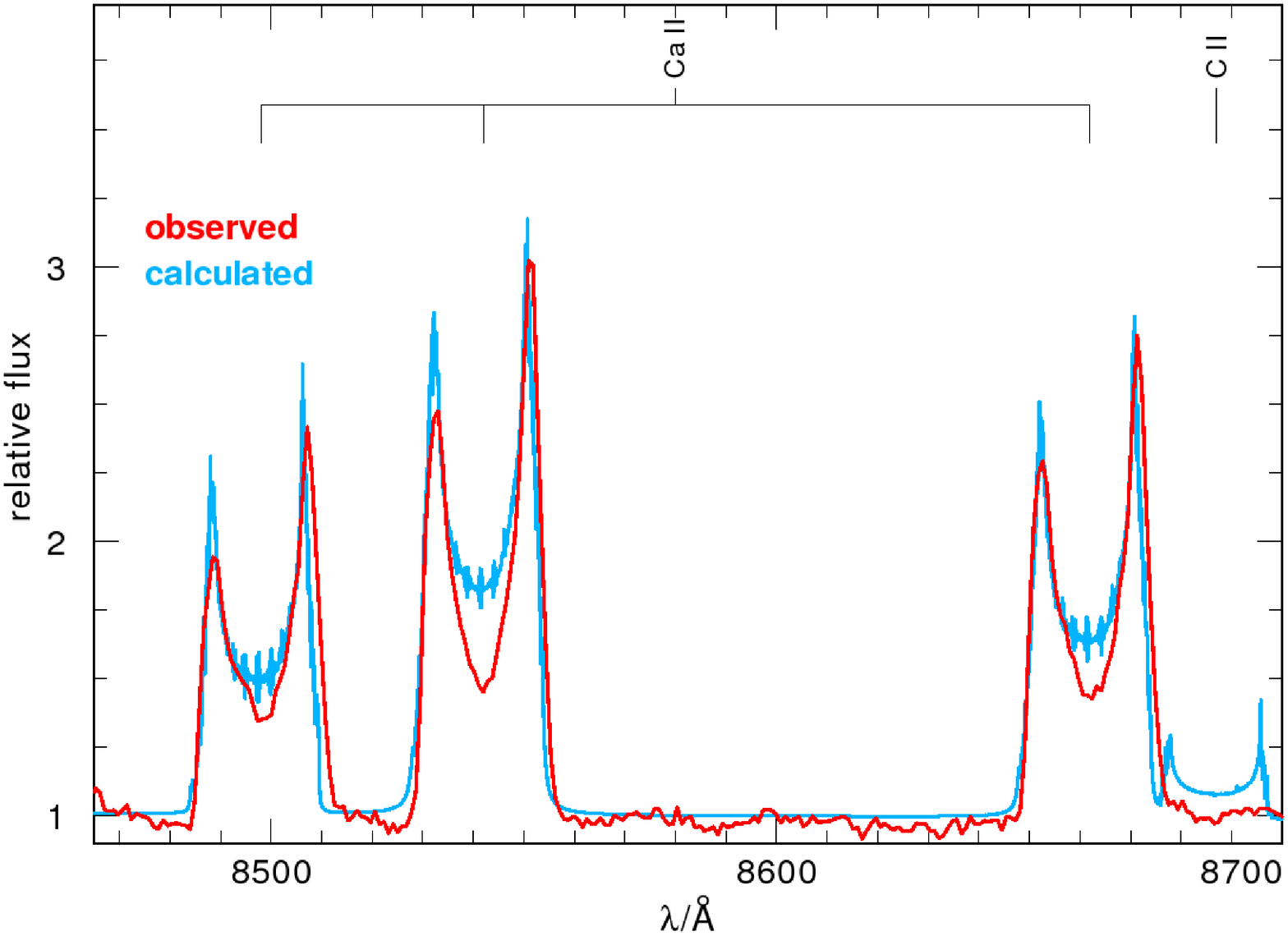}}
  \end{center}%
    \caption{Synthetic spectra for both geometries (shown in Fig.\,\ref{fig:6}), in comparison with the observation. The line asymmetry as well as the flat continuum between the lines are fitted well for an inclination of $i={76}^{\circ}$. The marked C\,\textsc{ii} feature seen to the far right of the modeled spectrum is actually a yet unsplit doublet line.}%
    \label{fig:8}%
}
\end{figure}

Calculated with the model atoms given in Tab.\,\ref{tab:2} and a set of parameters (Tab.\,\ref{tab:3}), the spectra of the two geometry maps are shown in Fig.\,\ref{fig:8}.\par
\begin{table}[b]%
  \centering%
  \begin{tabular}{l*{6}{r}}\hline
    Ions& H\,{\sc i-ii}& C\,{\sc i-iv}& O\,{\sc ii-iv}& Mg\,{\sc i-iii}& Si\,{\sc i-iv}& Ca\,{\sc i-iv}\\
    NLTE levels& $11$& $45$& $32$& $57$& $40$& $38$\\
    Line transitions& $45$& $88$& $65$& $104$& $49$& $64$\\
    Abundance (\%, in mass frac.)& $10^{-8}$& $4.6$& $65.5$& $13.5$& $15.1$& $1.3$\\\hline
  \end{tabular}%
  \caption{Model atoms and element abundances}%
  \label{tab:2}%
\end{table}
\begin{table}%
  \centering%
  \begin{tabular}{llr}\hline
    Ring radii range&$R/R_{\mathrm{WD}}$&$2\,-\,136$\\
    Disk surface mass density&$\Sigma/(\mathrm{g}/{\mathrm{cm}}^{2})$&$0.3$\\
    Disk temperature range&$T_{\mathrm{eff}}/\mathrm{K}$&$6700\,-\,5600$\\\hline
  \end{tabular}%
  \caption{\textsc{AcDc} integration parameters}%
  \label{tab:3}%
\end{table}%
As a first result of our analysis, we had to increase the inner radius of the disk to at least $R_{\mathrm{i}}{>}58\,R_{\mathrm{WD}}$ in order to suppress broad line wings and to fit the flat continuum between the triplet components as well as the line's steep profiles. The latest result by \citet{2010arXiv1007.2023M} being $R_{\mathrm{i}}=40\pm3\,R_{\mathrm{WD}}$ is only slightly smaller than our value. 
As matching the asymmetric line profile was the main goal of this work, we were able to achieve this well for both geometries, although the relative strengths of the red/violet parts of each line seem to be generally estimated too low in the case of the closed disk structure.\par
Of course the effect of the geometrical asymmetry strongly depends on the orientation towards the observer. With the rotation of the asymmetrical disk around the central object, the line profile should vary. \citet[priv.~comm.]{2008MNRAS.391L.103G} did, in fact, report on a change in the line profile in the order of years for at least two of the known five objects. But the predicted variability timescale of our model seems to be much shorter (order of hours) than the observations do suggest.\par
On the other hand, there might also be a long-time variability as the \textsc{fargo} simulations suggest a disk evolution through different geometries in the time of hundreds of orbits.
%
%
\begin{theacknowledgments}
  We thank Boris G\"ansicke for sending us his SDSS\,J1228+1040 spectrum in electronic form and for useful discussions. Also we like to thank Tobias M\"uller from the Computational Physics group in the Institute for Astronomy and Astrophysics of the University of T\"ubingen for providing us with the \textsc{fargo} code and the hydrodynamic simulations. T.R. is supported by the German Aerospace Center (DLR) under grant 05\,OR\,0806.
\end{theacknowledgments}
%
%

%
\end{document}